\documentclass[submission,copyright]{eptcs}
\providecommand{\event}{VPT 2016}

\usepackage{boxedminipage}
\usepackage{coqdoc}
\usepackage{graphicx}
\usepackage{url}

\usepackage{listings}
\newenvironment{lemme}{\begin{center}\begin{boxedminipage}{\textwidth}}{\end{boxedminipage}\end{center}}

\title{Renaming Global Variables in C\\ Mechanically Proved Correct}

\author{Julien Cohen
\institute{Universit\'e de Nantes, France}\\
 \email{Julien.Cohen@univ-nantes.fr}}

\def\authorrunning{J. Cohen}
\def\titlerunning{Renaming Global Variables in C Mechanically Proved Correct}

\begin{document}

\maketitle

\begin{abstract}
 Most integrated development environments are shipped with refactoring tools.
 However, their refactoring operations are often known to be unreliable.
As a consequence, developers have to test their code after applying an automatic refactoring.
In this article, we consider a refactoring operation (renaming of global variables in C),
 and we prove that its core implementation preserves the set of possible behaviors of transformed programs.
That proof of correctness relies on the operational semantics of C provided by CompCert C in Coq.
\end{abstract}

\section{Introduction}%%%%%%%%%%%%%%%%%%%%%%%%%%%%%%%%%%%%%%%%%%%%%%%%%%%%%%%%%%%%%%

\subsection{Refactoring tools are unreliable}

%\paragraph{The problem: Refactoring tools are not reliable.}
Designing refactoring tools is a complex task because of the complexity of the underlying programming languages.
As a result, probably all refactoring tools suffer from small bugs, that occur in rare cases, but that make those tools unreliable.
Many programmers have faced situations where the refactoring tool changed their program in an unexpected way.
Recent works have found tens of bugs in several refactoring tools~\cite{Mongiovi:2011, Soares:2012} by systematic testing.
This situation prevents programmers to trust their tools.

Some refactoring tools, such as the Haskell
Refactorer~\cite{HaRe},  are rigorously designed and
developed, but they are finally not free of bugs.\footnote{
The author reported several bugs in the Haskell Refactorer (2010, 2011).}

\subsection{Proving properties of refactoring tool operations}
Testing and proving are complementary approaches to software
validation.
Although the need for safe refactoring tools is
recognized~\cite{Schafer:2009, Brant:2015}, few efforts have been done to prove the correctness of such tools. 

In many research papers on refactoring, correctness is discussed informally.
Some papers give formal arguments,
but they either do not cover completely the preservation of
behaviors (for
instance:~\cite{CorrectRefactoringConcurrentJava}) or they
do not cover a complete programming language (for
instance:~\cite{HaRe-Formalisation}).
Also, existing mechanized proofs cover only theoretical languages (for instance:~\cite{Sultana-Thompson2008, Sultana-Thompson-2}).
There is a gap between the tools that are available for mainstream languages and the tools that are reliable.

\paragraph{Formalized refactoring in CompCert C.}
Our goal in this paper is to make a step towards a fully formalized refactoring tool for an industrial language. 
We choose to work on the language C because it has a mechanized formalization: CompCert C~\cite{Blazy-Leroy-Clight-09, Compcert-C}.
We present in section~\ref{sec:refactoring-compcert} how we use CompCert~C to build a refactoring tool.
Then we focus on a refactoring operation: renaming global variables. 
Renaming  seems  to be inoffensive at first sight, but we will see that there are some pitfalls to be avoided (Sec.~\ref{sec:renaming:spec}).
We show how our implementation handles several situations of shadowing (Sec.~\ref{sec-description-operation}) and that it preserves the behavior of programs (Sec.~\ref{sec-preservation-result}).
We also give a sufficient precondition for the operation (Sec.~\ref{sec:precondition}).

All the properties we give are proved in Coq. 
The full code with the proofs is available from the author or the project web-page.

\section{Refactoring in CompCert C}
\label{sec:refactoring-compcert}

To be able to prove properties on the behavior of transformed programs,
we need a formal definition of the semantics of programs.
We rely on the semantics of C programs formalized in Coq by CompCert C, which takes into account a subset of ISO C 99 larger than MISRA C 2004\footnote{We have used CompCert C version 2.4. See \url{http://compcert.inria.fr/compcert-C.html\#subset} for the list of supported features.
}.

That semantics (module \texttt{Csem} in CompCert) is defined on abstract syntax trees (AST). For this reason, we focus on AST transformations in this paper.
This allows to verify the logic of the transformation, but it also has some limits as described below (Sec.~\ref{sec:limits}).

\subsection{CompCert C syntax and semantics}

\paragraph{Abstract Syntax Trees.}

Identifiers (\texttt{AST.ident}) are represented by integers \linebreak(\texttt{BinNums.positive}). A map from textual identifiers in the original program to integer identifiers is maintained during parsing (function \texttt{intern\_string} defined in the OCaml module \texttt{Camlcoq}).

Programs (\texttt{AST.program}) are represented by a list of definitions (or declarations) of global variables and functions. 
A definition is represented by a pair $(i,d)$ where $i$ is the defined identifier and $d$ is the content of that definition.
Global variable definitions are composed of a type and an initialization (\texttt{AST.globdef}).
Function definitions are composed of a return type, a list of parameters, a list of local variables and a body (\texttt{Csyntax.fundef}). 
Function bodies are represented by statements (\texttt{Csyntax.statement}).
Statement syntax trees follow a 13 cases grammar, and rely on syntax trees of expressions (\texttt{Csyntax.expr}) that follow a 22 cases grammar.
In that grammar for expressions, we find the construction \texttt{Evar (x:ident) (ty:type)} to represent local or global variables or function names.

\paragraph{Formal semantics.}%%%%%%%%%%%%%%%%%%%%%%%%%%%%%%%%%%%

The semantics for C programs given in CompCert is based on small-step transitions (relation \texttt{Csem.step}). That relation respects the non-determinism of C programs: several transitions may be allowed from a given state.

The transitive closure of \texttt{step} is used to define the relation \texttt{Behaviors.program\_behaves} between programs and their possible behaviors. Behaviors are represented by the following datatype (in module \texttt{Behaviors}):

\medskip

\begin{center}
\begin{minipage}{12cm}
\begin{verbatim}
Inductive program_behavior: Type :=
  | Terminates: trace -> int -> program_behavior
  | Diverges: trace -> program_behavior
  | Reacts: traceinf -> program_behavior
  | Goes_wrong: trace -> program_behavior.
\end{verbatim}
\end{minipage}
\end{center}

\medskip

Those behaviors embed  \emph{traces} which are finite or infinite lists of observable events (\texttt{trace} and \texttt{traceinf}).

\begin{figure}[h]
\includegraphics[width=\textwidth]{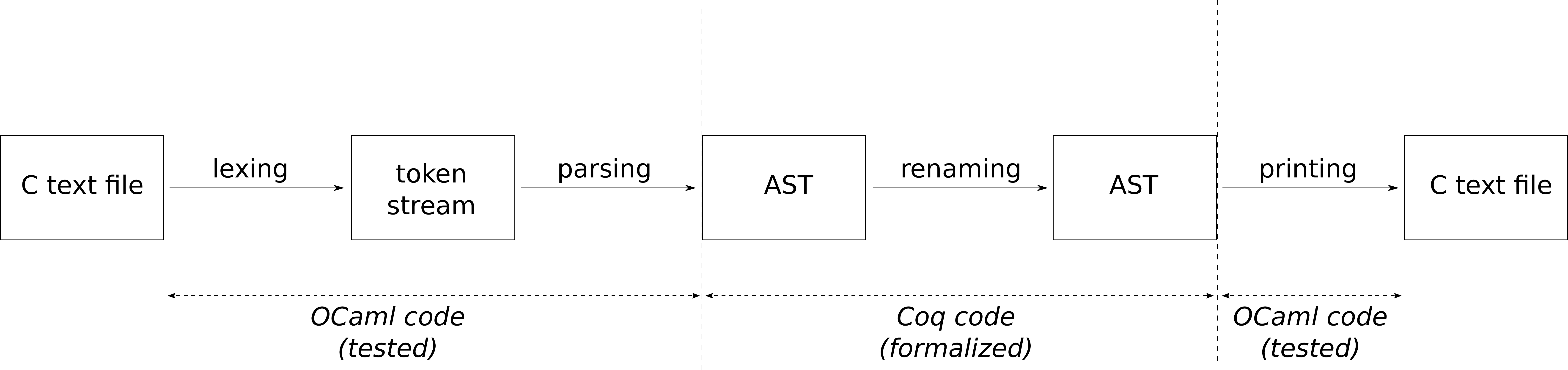}
\caption{Data-flow of the tool.}
\label{fig:data-flow}
\end{figure}

\subsection{Program Transformations}%%%%%%%%%%%%%%%%%%%%%%%%%%%%%%%%%%%%%%%%%
Our prototype follows the data-flow given in Fig~\ref{fig:data-flow}.
We use CompCert lexer, parser and pretty-printer.
The core refactoring is performed on parsed syntax trees.
Then the transformed AST are pretty-printed to recover a source file.
Most refactoring tools transform simultaneously the syntax tree and the token stream in order to recover the layout in the source file, but this is out of the scope of this paper.

A program transformation may fail and return an error (\texttt{Error} constructor with a message) or succeed and return the transformed AST (\texttt{OK} constructor with the resulting program, see the module \texttt{Errors} of CompCert).

\subsection{Behavior preservation}%%%%%%%%%%%%%%%%%%%%%%%%%%%%%%%%%%%%%%%%%%%%

\paragraph{Strict preservation.}
Considering a transformation that successfully transforms a program~\texttt{p} into a program~\texttt{t\_p}, the external behavior is \emph{strictly} preserved when \texttt{p} and \texttt{t\_p} have the same set of possible behaviors (type \texttt{program\_behavior}) with respect to the relation~\texttt{Behaviors.program\_behaves} (there is as \emph{bisimulation} between \texttt{p} and \texttt{t\_p}). 

\paragraph{Relaxed preservation.} \label{sec:relaxed-preservation} Some refactoring operations may not perform a strict behavior preservation, while still useful for users.
 In that case, we can precisely tell how the possible behaviors are modified.
 To do that, we exhibit a relation between the set of
 possible behaviors of \texttt{p} and the set of possible
 behaviors of \texttt{t\_p}.
For the renaming of global variables, we show in
Sec.~\ref{problem:extfun} that the set of behaviors is preserved up to
renaming in the traces.
We give a second example of relaxed preservation below.

\paragraph{Example.}
Consider the two programs below and a refactoring operation \emph{Extract variable} that would transform the program on the left side into the one on the right side (when applied two times).
The two programs do not have the same set of possible traces: the first one can print \texttt{AB} and \texttt{BA} whereas the second can only print \texttt{AB}.

\medskip

\noindent \hfill
\begin{minipage}{6.5cm}
\begin{lstlisting}
int main(){


  return printf("A") + printf("B");
  }
\end{lstlisting}
\end{minipage}
\hfill
\begin{minipage}{6.5cm}
\begin{lstlisting}
int main(){
  int r1 = printf("A") ;
  int r2 = printf("B") ;
  return r1 + r2 ;
  }
\end{lstlisting}
\end{minipage}
\hfill
~

\medskip

Here, the provider of that refactoring operation can ensure to the user that the set of behaviors of the resulting program is \emph{included in} (but not equal to) the set of behaviors of the original one (\emph{backward simulation}).

Note that strict preservation and relaxed preservation deal only with external behaviors (termination, returned results, traces), as is generally accepted for refactoring operations.

\subsection{Limits of the approach}%%%%%%%%%%%%%%%%%%%%%%%%%%%%%%%%%%%%%%%%%
\label{sec:limits}

Working on syntax trees implies the following limits:

\begin{itemize}

\item(Pre-processing.) We do not take pre-processing into account. We do not reconstruct pre-processor directives.

\item(Lexing and pretty-printing.) We do not preserve layout and comments.

\item(Block variables.) In CompCert syntax trees, block variables are encoded by function local variables. 
It makes as if all local variables of a function were declared at the beginning of its body.
Since we consider only syntax trees, we have no way to consider different maskings in different blocks of functions. 
Because of that, our renaming operation detects some capture situations in the AST that do not occur in the text source file, and fails to perform the renaming whereas it would be legal. The following program is an example of that situation (rename $x$ into $y$).

\lstset{keywords={x, y, main}}

\begin{center}
\begin{minipage}{8cm}
\begin{lstlisting}
int x = 1 ;

void main(void){
  x++ ;
  {
    int y = 1 ;
    y++ ;
  }          
} 
      /* Renaming x into y is correct */
      /* in this program.             */
\end{lstlisting}
\end{minipage}
\end{center}

\label{limit:block-variables}

That problem does not affect the correctness of the refactoring operation but it prevents its complete\-ness.

\item Some parts of the tool (parser, pretty-printer...) cannot be proven correct in the the same framework and must be tested.

\item We cannot perform renaming in programs that contain some syntax errors.

\end{itemize}

\section{Renaming Global Variables}%%%%%%%%%%%%%%%%%%%%%%%%%%%%%%%%%%%%%%%%%%%%%%%
\label{sec:renaming}

We now focus on a refactoring operation that renames global variables.
Renaming is one of the refactoring operations programmers use most.
Because of many shadowing and capture situations, renaming is often considered difficult.

\subsection{Analysis of the problem}%%%%%%%%%%%%%%%%%%%%%%%%%
\label{sec:renaming:spec}

\subsubsection{Dealing with shadowing} \label{sec:shadowing-spec}
The interesting part of renaming variables is dealing with shadowing (a local variable \emph{shadows} a global variable when they have the same name). 
To be able to preserve the behavior, we cannot create captures. In the following program, the renaming of $x$ into $y$  must fail:

\lstset{keywords={x, y, f}}

\begin{center}
\begin{minipage}{8.2cm}
\begin{lstlisting}[linewidth=8.2cm]
int x ;
int f(int y){
  return y + x ;
}
   /* Renaming x into y is NOT correct */
   /* in this program (capture).       */
\end{lstlisting}
\end{minipage}
\end{center}

However, we want to be able to introduce shadowings as long as they do not produce a capture.
For instance, renaming $x$ into $y$ in the following program introduces a new shadowing, but is correct:

\begin{center}
\begin{minipage}{8.2cm}
\begin{lstlisting}[linewidth=8.2cm]
int x ;
int f(int y){
  return y + 1 ;
}
      /* Renaming x into y is correct */
      /* in this program.             */
\end{lstlisting}
\end{minipage}
\end{center}

\subsubsection{Volatile Variables}%%%

We do not rename volatile variables because they are designed to be shared with the outside world.

\subsubsection{External Functions}%%%%%%%%%%%%%%%
\label{problem:extfun}

\paragraph{The problem.}

In C, linked libraries have access to global variables of the program as in the following example (consider the source code of the library is not available):

\medskip

\lstset{keywords={a, main, blackbox}}

\hfill\begin{minipage}{9cm}
\begin{lstlisting}[linewidth=9cm]
/* The library (shipped without source code) */

extern int a ;  /* This is a declaration */
                /* (not a definition).   */

void blackbox(){ a++ ; }
\end{lstlisting}
\end{minipage}

\noindent
\begin{minipage}{10cm}
\begin{lstlisting}[linewidth=10cm]
/* The main program */

void blackbox() ; /* External function declaration. */

int a = 0 ;       /* Global variable definition.    */

int main(){
  blackbox() ;
  return a ;
}
\end{lstlisting}
\end{minipage}

We generally cannot or do not want to propagate the renaming in libraries.
Here, renaming $a$ into $b$ only in the main program would change the behavior (it introduces an error).\footnote{For instance, Eclipse  performs the faulty renaming in the source file without warning (tested with Eclipse 4.5.1).}

\paragraph{Sufficient Condition as an Hypothesis.} 
Analysis of the compiled code of libraries is out of the scope of our prototype. 
So we have to assume that the renamed variable is not accessed from external code.
We also assume that the new name is not used for a global variable or function in the library.
Those assumptions are formalized by a predicate, \texttt{extcall\_additional\_properties} (see module \texttt{ExtCall} in the distributed source code). 
 That predicate is used as a precondition for our result on behavior preservation, (hypothesis EXT1 in Fig.~\ref{fig:main-result}).
The same assumption is made for inline assembly code that we do not analyze (hypothesis EXT2 in Fig.~\ref{fig:main-result}).
Those assumptions are made only for the two names involved in the renaming.

\subsection{Implementation of the transformation}%%%%%%%%%%%%%%%%%%%%%%%
\label{sec-description-operation}
We now describe our implementation.
In the following, we consider $x$ is the name to be changed, $y$ is the new name, and $x\not = y$.

\paragraph{Pre-operations.}

Before calling the transformation defined in Coq we perform the following operations (coded in OCaml): \begin{enumerate}

\item Check that $y$ is not a C keyword. This cannot be done in the Coq part because identifiers are represented by numbers in syntax trees. 

\item Trigger parsing. We use the CompCert parser, but we must not use the feature of the \mbox{CompCert} compiler that changes the names of the variables to have a unique name for each variable. Otherwise, we would have no way to detect shadowings in the AST.
\end{enumerate}

\paragraph{Top-level checks and transformation.}
When the AST is available from parsing, the transformation defined in Coq (function \texttt{rename\_globvar\_hard} in module \texttt{Programs}) 
makes the following verifications:

\begin{enumerate}

\item Check that $x$ and $y$ are different from $main$.

\item Check that $x$ is declared as a global variable.

\end{enumerate}

Then it triggers the transformation of all definitions. For each definition, the function \texttt{rename2} in module \texttt{Definitions.Def}
makes the following verifications:

\begin{enumerate}

\item If it defines/declares $x$, we check that: \begin{enumerate}

\item the definition is not for a function but a global variable ;
\item $x$ does not appear in its initialization ;
\item $y$ does not appear in the initialization ;
\item the variable is not volatile.

\end{enumerate}

Then we change the name of the definition into $y$.

\item Check that it does not define/declare $y$ .

\item In other cases, propagate the renaming in the content of the definition: function or global variable initialization.

\end{enumerate}
We describe next how the renaming in functions is performed.

\paragraph{Renaming in functions.}

To propagate a renaming in a function $f$, we first check if $f$ binds $x$ and $y$ by the means of a parameter or a local variable (see \texttt{propagate\_change\_ident} below\footnote{The notation \texttt{do s <- E1 ; return E2} is a shortcut for \texttt{match E1 with OK s => E2 | Error e => Error e end}. See the monad \texttt{Error} defined in CompCert, which is used to represent computations that can fail. The function \texttt{dec\_binds} checks if a variable is bound in a function by the means of a formal parameter or a local variable.}).

\medskip

\noindent
\begin{boxedminipage}{\columnwidth}

\begin{coqdoccode}

\coqdocnoindent
\coqdockw{Definition} \coqdocvar{propagate\_change\_ident} (\coqdocvar{x}:\coqdocvar{AST.ident}) (\coqdocvar{y}:\coqdocvar{AST.ident}) (\coqdocvar{f}:\coqdocvar{Csyntax.function}) :=\coqdoceol
\coqdocnoindent
\coqdoceol
\coqdocindent{3.00em}
\coqdockw{if} \coqdocvar{dec\_binds} \coqdocvar{x} \coqdocvar{f} \coqdoceol
\coqdocindent{3.00em}
\coqdockw{then} \coqdoceol
\coqdocindent{4.00em}
\coqdockw{if} \coqdocvar{dec\_binds} \coqdocvar{y} \coqdocvar{f} \coqdoceol
\coqdocindent{4.00em}
\coqdockw{then} \coqdocvar{OK} \coqdocvar{f} \coqdoceol
\coqdocindent{4.00em}
\coqdockw{else} \coqdoceol
\coqdocindent{5.00em}
\coqdockw{if} \coqdocvar{dec\_appears\_statement} \coqdocvar{y} (\coqdocvar{fn\_body} \coqdocvar{f})\coqdoceol
\coqdocindent{5.00em}
\coqdockw{then} \coqdocvar{Error} (\coqdocvar{msg} "Replacing identifier occurring in function.")\coqdoceol
\coqdocindent{5.00em}
\coqdockw{else} \coqdocvar{OK} \coqdocvar{f}\coqdoceol
\coqdocnoindent
\coqdoceol
\coqdocindent{3.00em}
\coqdockw{else} \coqdoceol
\coqdocindent{4.00em}
\coqdockw{if} \coqdocvar{dec\_binds} \coqdocvar{y} \coqdocvar{f} \coqdoceol
\coqdocindent{4.00em}
\coqdockw{then} \coqdoceol
\coqdocindent{5.00em}
\coqdockw{if} \coqdocvar{dec\_appears\_statement} \coqdocvar{x} (\coqdocvar{fn\_body} \coqdocvar{f})\coqdoceol
\coqdocindent{5.00em}
\coqdockw{then} \coqdocvar{Error} (\coqdocvar{msg} "This renaming would introduce an undesired shadowing.") \coqdoceol
\coqdocindent{5.00em}
\coqdockw{else} \coqdocvar{OK} \coqdocvar{f}\coqdoceol
\coqdocindent{4.00em}
\coqdockw{else} \coqdocvar{force\_body} \coqdocvar{x} \coqdocvar{y} \coqdocvar{f}.\coqdoceol

\coqdocemptyline

\medskip
\hrule
\medskip

\coqdocemptyline
\coqdocnoindent
\coqdockw{Definition} \coqdocvar{force\_body} (\coqdocvar{x}:\coqdocvar{AST.ident}) (\coqdocvar{y}:\coqdocvar{AST.ident}) (\coqdocvar{f}:\coqdocvar{Csyntax.function}) := \coqdoceol
\coqdocnoindent
\coqdoceol
\coqdocindent{1.00em}
\coqdoctac{do}  \coqdocvar{s} \ensuremath{\leftarrow} \coqdocvar{Statements.change\_ident} \coqdocvar{x} \coqdocvar{y} (\coqdocvar{fn\_body} \coqdocvar{f}) ;\coqdoceol
\coqdocindent{1.00em}
\coqdocvar{OK} (\coqdocvar{mkfunction} (\coqdocvar{fn\_return} \coqdocvar{f}) (\coqdocvar{fn\_callconv} \coqdocvar{f}) (\coqdocvar{fn\_params} \coqdocvar{f}) (\coqdocvar{fn\_vars} \coqdocvar{f}) \coqdocvar{s}).\coqdoceol
\coqdocemptyline

\end{coqdoccode}

\end{boxedminipage}

\medskip

Four different situations may occur:

\begin{itemize}

\item $f$ does not bind $x$ / $f$  does not bind $y$: Rename $x$ into $y$ in the body of $f$ (function \texttt{force\_body}). This will report a failure if $y$ is encountered,  report a success otherwise.

\item $f$ binds $x$ / $f$  binds $y$: Do not change the body of $f$ because of a shadowing (success).

\item $f$ does not bind $x$ / $f$  binds $y$:
  If $x$ appears in the body of $f$, report a failure to avoid a capture. Else do not change the body of $f$ (success).

\item $f$ binds $x$ / $f$  does not bind $y$:
  Do not change the body of $f$ because of a shadowing, but check that $y$ does not occur in the body of $f$. Otherwise, we would transform an ill-formed program into a well-formed program, as for the following program:\footnote{ \label{footnote:bug-whole-tomato}
 For instance, for that program, some refactoring tools (such as Eclipse and Visual Assist / Whole Tomato Software for Visual Studio) will perform the renaming of the global $x$ into $y$, yielding a valid program from an invalid one, which obviously changes the semantics.}
\end{itemize}

\lstset{keywords={x, y, f}}

\begin{center}
\begin{minipage}{9cm}
\begin{lstlisting}
int x;

int f(int x){
  return y ;    /* This instance of y is free. */
}
\end{lstlisting}
\end{minipage}
\end{center}

\paragraph{Renaming in statements and expressions.}
Function bodies are represented by statements.
To rename a statement, we just propagate the renaming to the leaves of the syntax tree that contain occurrences of variables (constructor \texttt{Evar} of expressions).
A failure is reported if $y$ is encountered as a variable name (labels are not checked).

\medskip

\noindent
\begin{boxedminipage}{\columnwidth}
\smallskip
\begin{coqdoccode}
\coqdocnoindent
\coqdockw{Definition} \coqdocvar{change\_ident\_untyped} (\coqdocvar{x} : \coqdocvar{AST.ident}) (\coqdocvar{y}:\coqdocvar{AST.ident}) \coqdocvar{i} :=\coqdoceol
\coqdocindent{5.00em}
\coqdockw{if} \coqdocvar{AST.ident\_eq} \coqdocvar{x} \coqdocvar{i}\coqdoceol
\coqdocindent{5.00em}
\coqdockw{then} \coqdocvar{OK} \coqdocvar{y}          \coqdoceol
\coqdocindent{5.00em}
\coqdockw{else}\coqdoceol
\coqdocindent{6.00em}
\coqdockw{if} \coqdocvar{AST.ident\_eq} \coqdocvar{i} \coqdocvar{y}\coqdoceol
\coqdocindent{6.00em}
\coqdockw{then} \coqdocvar{Error} (\coqdocvar{msg} "replacing identifier already occurs")\coqdoceol
\coqdocindent{6.00em}
\coqdockw{else} \coqdocvar{OK} \coqdocvar{i} .
\coqdoceol
\coqdocemptyline
\end{coqdoccode}
\smallskip
\end{boxedminipage}

\medskip

\paragraph{Optimized syntax trees.}
\label{sec:RG}

 The datatype for statements has some constructors to represent optimized forms of accesses to global variables.
We report a failure when those constructors are encountered while renaming a statement, but that case never occurs since we deal with not-optimized syntax trees (those optimizations take place at compilation, not at parsing).
Such cases are excluded with the hypothesis RG of our result of correctness (Fig.~\ref{fig:main-result}).

\subsection{Trace and behavior correspondence}
\label{sec:transformation-traces}

In CompCert, traces include some references to global variables (read and write accesses).
Those references help to prove the preservation of behaviors by compilation steps of the CompCert compiler. 
They are not really part of the external
behavior: they cannot be observed externally when the variables are
not volatile and when libraries do not access them.

The presence of those references makes it impossible to preserve strictly the traces when you change some global variable names.
So we want to characterize precisely those changes to be able to tell if we accept them (relaxed behavior preservation as explained in Sec~\ref{sec:relaxed-preservation}).

First, we build two functions that perform renaming of global variables in finite traces \linebreak (\texttt{Events.rename\_in\_trace}) and in infinite traces (\texttt{TraceInf.rename\_traceinf}).
Then we use those functions to build a function (\texttt{Behaviors.rename\_globvar}) that renames global variables occurring in behaviors.
That function is used to prove relaxed preservation of behaviors (Sec~\ref{sec-preservation-result}, Fig~\ref{fig:main-result}).

\begin{figure}[!h]

\noindent
\begin{lemme}

% Variables x y
\begin{coqdoccode}
\coqdocnoindent
\coqdockw{Variable} \coqdocvar{x} : \coqdocvar{AST.ident}.\coqdoceol
\coqdocnoindent
\coqdockw{Variable} \coqdocvar{y} : \coqdocvar{AST.ident}.\coqdoceol
\coqdocemptyline
\end{coqdoccode}

% Fonctions externes
\begin{coqdoccode}
 \coqdocemptyline
\coqdocnoindent
\coqdockw{Hypothesis} \coqdocvar{EXT1} :\coqdoceol
\coqdocindent{1.00em}
\coqdockw{\ensuremath{\forall}} \coqdocvar{name} \coqdocvar{sig},\coqdoceol
\coqdocindent{2.00em}
\coqdocvar{ExtCall.extcall\_additional\_properties} (\coqdocvar{Events.external\_functions\_sem} \coqdocvar{name} \coqdocvar{sig}) \coqdocvar{x} \coqdocvar{y}.\coqdoceol
\coqdocemptyline
\coqdocnoindent
\coqdockw{Hypothesis} \coqdocvar{EXT2} :\coqdoceol
\coqdocindent{1.00em}
\coqdockw{\ensuremath{\forall}} \coqdocvar{text},\coqdoceol
\coqdocindent{2.00em}
\coqdocvar{ExtCall.extcall\_additional\_properties} (\coqdocvar{Events.inline\_assembly\_sem} \coqdocvar{text}) \coqdocvar{x} \coqdocvar{y}.\coqdoceol
\coqdocemptyline
\end{coqdoccode}

% Programme p et contrainte RawGlobals
 \begin{coqdoccode}
\coqdocemptyline
\coqdocnoindent
\coqdockw{Variable} \coqdocvar{p} : \coqdocvar{program}.\coqdoceol
\coqdocemptyline
\coqdocnoindent
\coqdockw{Hypothesis} \coqdocvar{RG} :  \coqdocvar{RawGlobals.rawglobals} \coqdocvar{p} .\coqdoceol
\coqdocemptyline
\end{coqdoccode}

% Résultat behavior\_preserved1
 \begin{coqdoccode}
\coqdocemptyline
\coqdocnoindent
\coqdockw{Theorem} \coqdocvar{behavior\_preserved\_1} : \coqdoceol
\coqdocindent{1.00em}
\coqdocvar{x} \ensuremath{\not=} \coqdocvar{y} \ensuremath{\rightarrow}\coqdoceol
\coqdocindent{1.00em}
\coqdockw{\ensuremath{\forall}} (\coqdocvar{t\_p}:\coqdocvar{program}) ,\coqdoceol
\coqdocindent{2.00em}
\coqdocvar{Programs.rename\_globvar\_hard} \coqdocvar{x} \coqdocvar{y} \coqdocvar{p} = \coqdocvar{OK} \coqdocvar{t\_p} \ensuremath{\rightarrow}\coqdoceol
\coqdocindent{2.00em}
\coqdockw{\ensuremath{\forall}} (\coqdocvar{b} : \coqdocvar{program\_behavior}),\coqdoceol
\coqdocindent{3.00em}
\coqdocvar{program\_behaves} (\coqdocvar{Csem.semantics} \coqdocvar{p}) \coqdocvar{b} \ensuremath{\rightarrow}\coqdoceol
\coqdocindent{3.00em}
\coqdoctac{\ensuremath{\exists}} \coqdocvar{t\_b}, \coqdoceol
\coqdocindent{4.00em}
( \coqdocvar{Behaviors.rename\_globvar} \coqdocvar{x} \coqdocvar{y} \coqdocvar{b} = \coqdocvar{OK} \coqdocvar{t\_b} \ensuremath{\land}\coqdoceol
\coqdocindent{5.00em}
\coqdocvar{program\_behaves} (\coqdocvar{Csem.semantics} \coqdocvar{t\_p}) \coqdocvar{t\_b} ).\coqdoceol
\end{coqdoccode}

\end{lemme}
\caption{Forward simulation}
\label{fig:main-result}
\end{figure}

\begin{figure}[!h]

\noindent
\begin{lemme}
\begin{coqdoccode}
\coqdocnoindent
\coqdockw{Theorem} \coqdocvar{behavior\_preserved\_3} : \coqdoceol
\coqdocindent{1.00em}
\coqdocvar{x}\ensuremath{\not=}\coqdocvar{y} \ensuremath{\rightarrow}\coqdoceol
\coqdocindent{1.00em}
\coqdockw{\ensuremath{\forall}} (\coqdocvar{t\_p}:\coqdocvar{program}),\coqdoceol
\coqdocindent{2.00em}
\coqdocvar{Programs.rename\_globvar\_hard} \coqdocvar{x} \coqdocvar{y} \coqdocvar{p} = \coqdocvar{OK} \coqdocvar{t\_p} \ensuremath{\rightarrow}\coqdoceol
\coqdocindent{2.00em}
\coqdockw{\ensuremath{\forall}} (\coqdocvar{t\_b} : \coqdocvar{program\_behavior}),    \coqdoceol
\coqdocindent{3.00em}
\coqdocvar{program\_behaves} (\coqdocvar{Csem.semantics} \coqdocvar{t\_p}) \coqdocvar{t\_b} \ensuremath{\rightarrow}\coqdoceol
\coqdocindent{3.00em}
\coqdoctac{\ensuremath{\exists}} \coqdocvar{b},\coqdoceol
\coqdocindent{4.00em}
( \coqdocvar{Behaviors.rename\_globvar} \coqdocvar{y} \coqdocvar{x} \coqdocvar{t\_b} = \coqdocvar{OK} \coqdocvar{b} \ensuremath{\land}\coqdoceol
\coqdocindent{5.00em}
\coqdocvar{program\_behaves} (\coqdocvar{Csem.semantics} \coqdocvar{p}) \coqdocvar{b} ).\coqdoceol
\coqdocemptyline
\end{coqdoccode}

\end{lemme}
\caption{Backward simulation (same hypoteses as for forward simulation)}
\label{fig:backard-simulation}
\end{figure}

\subsection{Properties}%%%%%%%%%%%%%%%%%%%%%%%%%%%%%%%%%%%

\subsubsection{Behavior preservation}%%%%%%%%%%%%%%%%%%%%%%%%%%%%%%%%%%%%%%%%%%%%%
\label{sec-preservation-result}

Under the conditions already discussed (hypotheses  \coqdocvar{EXT1}, \coqdocvar{EXT2} and \coqdocvar{RG} in Fig.~\ref{fig:main-result}),
and when the renaming operations succeeds,
 the transformed program has the same set of possible behaviors as the original program, up to renaming in the traces.
The equality of the two sets comes from a double inclusion. 
The first inclusion (forward simulation) is formalized by the theorem \mbox{\coqdocvar{behavior\_preserved\_1}} (Fig.~\ref{fig:main-result}, proved in Coq in module \texttt{Correctness}): 
if the original program can have a given behavior, then the transformed program can have the same behavior (up to renaming in its trace).

The second inclusion (backward simulation) is stated in the theorem \emph{behavior\_preserved\_3} (Fig.~\ref{fig:backard-simulation}): 
if the transformed program can have a behavior, then the original program can have the same behavior (up to renaming in its trace).

\begin{figure}[!h]

\noindent\begin{boxedminipage}{9cm}
\begin{coqdoccode}
\coqdocnoindent
\coqdockw{Lemma} \coqdocvar{step\_commut} :\coqdoceol
\coqdocindent{1.00em}
\coqdockw{\ensuremath{\forall}} \coqdocvar{x} \coqdocvar{y} \coqdocvar{ge} \coqdocvar{tr\_ge} \coqdocvar{st1} \coqdocvar{tr\_st1} \coqdocvar{tra} \coqdocvar{tr\_tra} \coqdocvar{st2} \coqdocvar{tr\_st2},\coqdoceol
\coqdocnoindent
\coqdoceol
\coqdocindent{2.00em}
( ... \texttt{(* same as EXT1 in Fig.~\ref{fig:main-result} *)} )
\ensuremath{\rightarrow}\coqdoceol
\coqdocindent{2.00em}
( ... \texttt{(* same as EXT2 in Fig.~\ref{fig:main-result} *)} ) \ensuremath{\rightarrow}\coqdoceol
\coqdocnoindent
\coqdoceol
\coqdocindent{2.00em}
\coqdocvar{GlobalEnv.rename\_globvar} \coqdocvar{x} \coqdocvar{y} \coqdocvar{ge} = \coqdocvar{OK} \coqdocvar{tr\_ge} \ensuremath{\rightarrow}\coqdoceol
\coqdocindent{2.00em}
\coqdocvar{State.change\_ident} \coqdocvar{x} \coqdocvar{y} \coqdocvar{st1} = \coqdocvar{OK} \coqdocvar{tr\_st1} \ensuremath{\rightarrow}\coqdoceol
\coqdocindent{2.00em}
\coqdocvar{State.change\_ident} \coqdocvar{x} \coqdocvar{y} \coqdocvar{st2} = \coqdocvar{OK} \coqdocvar{tr\_st2} \ensuremath{\rightarrow}\coqdoceol
\coqdocindent{2.00em}
\coqdocvar{rename\_in\_trace} \coqdocvar{x} \coqdocvar{y} \coqdocvar{tra} = \coqdocvar{OK} \coqdocvar{tr\_tra} \ensuremath{\rightarrow}\coqdoceol
\coqdocindent{2.00em}
\coqdocvar{step} \coqdocvar{ge} \coqdocvar{st1} \coqdocvar{tra} \coqdocvar{st2} \ensuremath{\rightarrow} \coqdoceol
\coqdocindent{2.00em}
\coqdocvar{RawGlobals.rawglobals\_state} \coqdocvar{st1} \ensuremath{\rightarrow}\coqdoceol
\coqdocindent{2.00em}
\coqdocvar{wf\_state} \coqdocvar{st1} \ensuremath{\rightarrow} \coqdoceol
\coqdocindent{2.00em}
\coqdocvar{step} \coqdocvar{tr\_ge} \coqdocvar{tr\_st1} \coqdocvar{tr\_tra} \coqdocvar{tr\_st2}.\coqdoceol
\coqdocemptyline
\end{coqdoccode}
\end{boxedminipage}
\hfill
\begin{minipage}{5cm}
\noindent
\includegraphics[width=5cm]{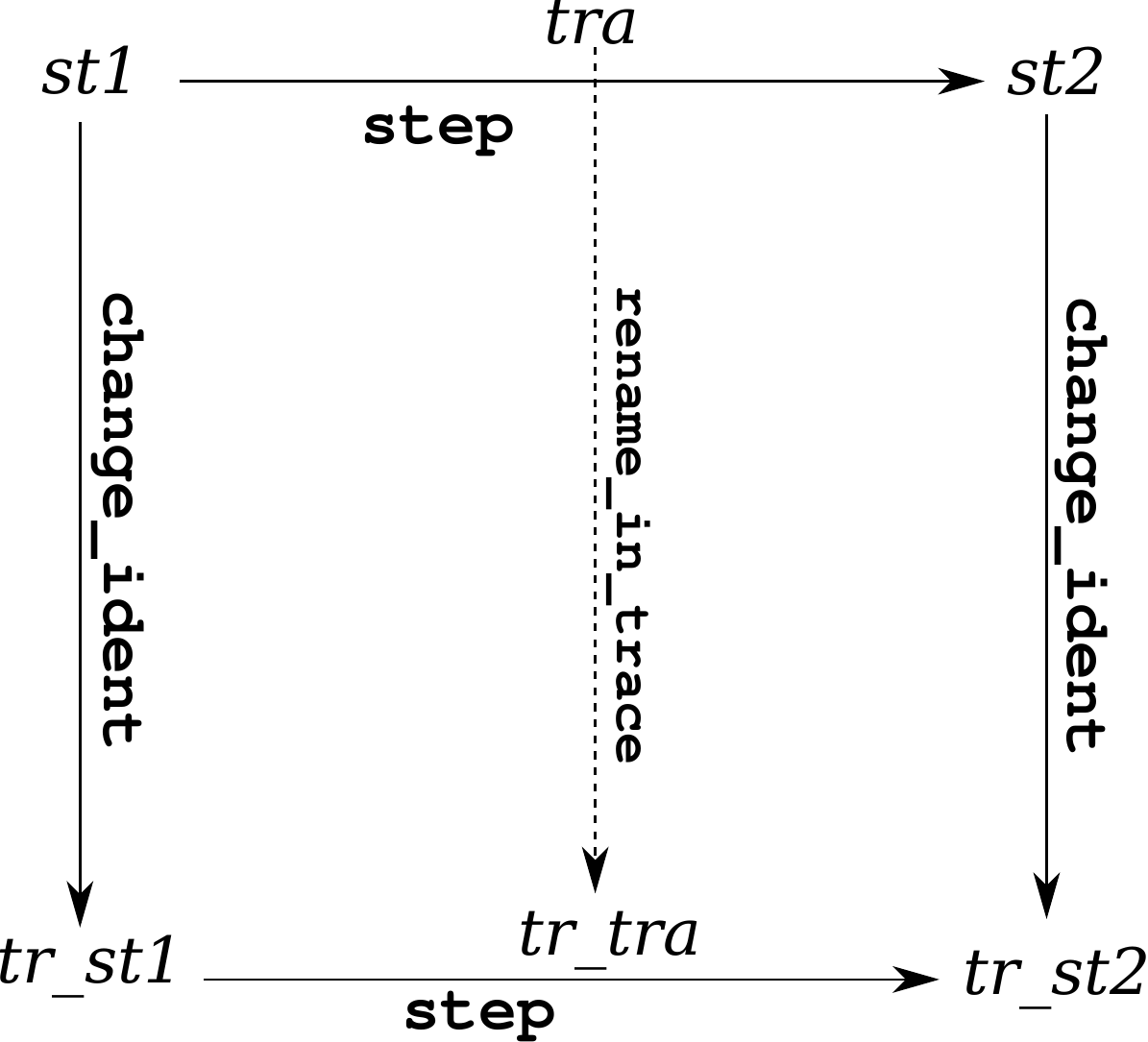}
\end{minipage}
\caption{Commutativity of renaming with transitions (relation \texttt{step})}
\label{fig:step-commut}
\end{figure}

\paragraph{Structure of the Proof.}%%%

The key point to prove the result of forward simulation above is the preservation of transitions (lemma \texttt{step\_commut} of Fig.~\ref{fig:step-commut}).

To prove this, we build a correspondence between states coming from the execution of the initial program with states in the execution of the transformed program (function \texttt{change\_ident} in module \texttt{State}).
That correspondence relies on correspondences we build on continuations, on contexts, and on global environments.
Some aspects of the proof are further discussed in App.~\ref{sec:proof-explaination}.

\subsubsection{Sufficient precondition (for the transformation to succeed)}%%%%%%%%%%%%%%%%%%%%%%%%%%%%%%%%%%%%%%%%%
\label{sec:precondition}

The result of correctness holds when the transformation succeeds (does not fail and return an error). 
Here we characterize the set of programs for which it succeeds to show it does not fail without a good reason.

To help to characterize problematic situations, we introduce the predicate \texttt{covers y x f} that says that renaming \texttt{x} into \texttt{y} in a function \texttt{f} would introduce a capture.

\begin{lemme}

\begin{coqdoccode}
\coqdocnoindent
\coqdockw{Definition} \coqdocvar{covers} \coqdocvar{y} \coqdocvar{x} \coqdocvar{f} := \coqdoceol
\coqdocindent{1.00em}
\coqdocvar{binds} \coqdocvar{y} \coqdocvar{f} \ensuremath{\land}  \ensuremath{\lnot}\coqdocvar{binds} \coqdocvar{x} \coqdocvar{f} \ensuremath{\land} \coqdocvar{appears\_statement} \coqdocvar{x} (\coqdocvar{fn\_body} \coqdocvar{f}).\coqdoceol
\coqdocemptyline\end{coqdoccode}
\end{lemme}

Note that this predicate is coherent with the two examples of program given in Sec.~\ref{sec:shadowing-spec}: $x$ is ``covered'' by $y$ in the first one while $x$ is not ``covered'' by $y$ in the second one.

The predicate \coqdocvar{no\_cover\_in\_prog} below says that no function of a program is subject to capture.

\begin{lemme}

\begin{coqdoccode}
\coqdocnoindent
\coqdockw{Definition} \coqdocvar{no\_cover\_in\_prog} \coqdocvar{x} \coqdocvar{y} (\coqdocvar{p} : \coqdocvar{program} \coqdocvar{Csyntax.fundef} \coqdocvar{Ctypes.type}) :=\coqdoceol
\coqdocindent{1.00em}
\coqdockw{\ensuremath{\forall}} (\coqdocvar{f} : \coqdocvar{function}) (\coqdocvar{i} : \coqdocvar{ident}),\coqdoceol
\coqdocindent{2.00em}
\coqdocvar{List.In} (\coqdocvar{i}, \coqdocvar{Gfun} (\coqdocvar{Csyntax.Internal} \coqdocvar{f})) (\coqdocvar{prog\_defs} \coqdocvar{p}) \ensuremath{\rightarrow}\coqdoceol
\coqdocindent{2.00em}
\ensuremath{\lnot}\coqdocvar{Fun.covers} \coqdocvar{y} \coqdocvar{x} \coqdocvar{f}.\coqdoceol
\coqdocemptyline
\end{coqdoccode}
\end{lemme}

The following result \coqdocvar{sufficient\_precondition} shows which conditions are sufficient for the transformation to succeed on a given program.
Some predicates that have not been explained here can be found in
the source code; they have the usual meaning the reader would probably expect.

\begin{lemme}
\begin{coqdoccode}
\coqdocnoindent
\coqdockw{Theorem} \coqdocvar{sufficient\_precondition} : \coqdoceol
\coqdocindent{1.00em}
\coqdockw{\ensuremath{\forall}} \coqdocvar{x} \coqdocvar{y} \coqdocvar{p},\coqdoceol
\coqdocindent{2.00em}
\coqdocvar{x} \ensuremath{\not=} \coqdocvar{y} \ensuremath{\rightarrow}\coqdoceol
\coqdocindent{2.00em}
\coqdocvar{x} \ensuremath{\not=} \coqdocvar{prog\_main} \coqdocvar{p} \ensuremath{\rightarrow}\coqdoceol
\coqdocindent{2.00em}
\coqdocvar{y} \ensuremath{\not=} \coqdocvar{prog\_main} \coqdocvar{p} \ensuremath{\rightarrow}\coqdoceol
\coqdocnoindent
\coqdoceol
\coqdocindent{2.00em}
\coqdocvar{RawGlobals.rawglobals} \coqdocvar{p} \ensuremath{\rightarrow}\coqdoceol
\coqdocnoindent
\coqdoceol
\coqdocindent{2.00em}
\coqdocvar{defines\_globvar} \coqdocvar{x} \coqdocvar{p} \ensuremath{\rightarrow}\coqdoceol
\coqdocindent{2.00em}
\ensuremath{\lnot}\coqdocvar{defines\_globvar} \coqdocvar{y} \coqdocvar{p} \ensuremath{\rightarrow}\coqdoceol
\coqdocindent{2.00em}
\ensuremath{\lnot}\coqdocvar{defines\_volatile\_globvar} \coqdocvar{x} \coqdocvar{p} \ensuremath{\rightarrow}\coqdoceol
\coqdocnoindent
\coqdoceol
\coqdocindent{2.00em}
\ensuremath{\lnot}\coqdocvar{defines\_func} \coqdocvar{x} \coqdocvar{p} \ensuremath{\rightarrow}\coqdoceol
\coqdocindent{2.00em}
\ensuremath{\lnot}\coqdocvar{defines\_func} \coqdocvar{y} \coqdocvar{p} \ensuremath{\rightarrow}\coqdoceol
\coqdocnoindent
\coqdoceol
\coqdocindent{2.00em}
\ensuremath{\lnot}\coqdocvar{appears\_free} \coqdocvar{y} \coqdocvar{p} \ensuremath{\rightarrow}\coqdoceol
\coqdocindent{2.00em}
\ensuremath{\lnot}\coqdocvar{appears\_free} \coqdocvar{x} \coqdocvar{p} \ensuremath{\rightarrow}\coqdoceol
\coqdocnoindent
\coqdoceol
\coqdocindent{2.00em}
\coqdocvar{no\_cover\_in\_prog} \coqdocvar{x} \coqdocvar{y} \coqdocvar{p} \ensuremath{\rightarrow}\coqdoceol
\coqdocnoindent
\coqdoceol
\coqdocindent{2.00em}
\coqdoctac{\ensuremath{\exists}} \coqdocvar{t\_p}, \coqdocvar{rename\_globvar\_hard} \coqdocvar{x} \coqdocvar{y} \coqdocvar{p} = \coqdocvar{OK} \coqdocvar{t\_p} .\coqdoceol
\end{coqdoccode}
\end{lemme}

Of course, this precondition applies on syntax trees, so we add that the parsing has to succeed for the transformation to succeed.

\subsubsection{Invertibility}

The transformation is invertible in the following meaning:

\begin{coqdoccode}
\coqdocemptyline
\coqdocnoindent
\coqdockw{Lemma} \coqdocvar{invertibility} :\coqdoceol
\coqdocindent{1.00em}
\coqdockw{\ensuremath{\forall}} \coqdocvar{x} \coqdocvar{y} \coqdocvar{p} \coqdocvar{r},\coqdoceol
\coqdocindent{2.00em}
\coqdocvar{rename\_globvar\_hard} \coqdocvar{x} \coqdocvar{y} \coqdocvar{p} = \coqdocvar{OK} \coqdocvar{r} \ensuremath{\rightarrow}\coqdoceol
\coqdocindent{2.00em}
\coqdocvar{rename\_globvar\_hard} \coqdocvar{y} \coqdocvar{x} \coqdocvar{r} = \coqdocvar{OK} \coqdocvar{p}.\coqdoceol
\coqdocemptyline
\end{coqdoccode}

\subsubsection{Alternate proof for backward simulation}

We have two very different proofs for the backward simulation (theorems \emph{behavior\_preserved\_2}, not included in the paper and \emph{behavior\_preserved\_3}, Fig.~\ref{fig:backard-simulation}). 
The first one relies on the same technique as the proof of forward simulation
whereas the second one comes for free from the invertibility of the transformation and forward simulation.

\section{Conclusion}%%%%%%%%%%%%%%%%%%%%%%%%%%%%

\subsection{Results}%%%%%%%%%%%%%%%%%%%%%%%%%%%%%%%%%%%%%%%%%%%%%%%%%%%%%%%%%%%%%%%%
We have built a refactoring tool whose logic part is
formally described and proved correct. 
Although it has some limits (layout, pre-processing directives, and separate compilation not well supported as discussed), our prototype produces C code that has the same semantics as the initial program, even when considering non-determinism.

\subsection{Related work}
Faced with the complexity of making proofs for large
languages such as C or Java, the community of refactoring
explored naturally the systematic testing of refactoring
tools~\cite{Mongiovi:2011, Soares:2012} and of refactored
programs~\cite{Mongiovi:2014, Ge:2014}.

Authors often give some properties of their refactoring operations, but generally in an informal way.
It is not surprising that the first formal works in that
domain, such as~\cite{HaRe-Formalisation}
and~\cite{Sultana-Thompson2008} (the latter is mechanized with Isabelle/HOL), were applied to functional
languages, where a long tradition of program transformations exist.
A few authors have a formal approach with imperative programs, but they focus on specific aspects of the transformation.
For instance, \cite{CorrectRefactoringConcurrentJava}
considers as an hypothesis that refactoring operations
preserve the behavior for sequential executions in order to prove the
preservation of the behavior for concurrent executions.
To the best of our knowledge, our work is the first to prove formally the behavior preservation for an industrial language.

\subsection{Open questions and Future Work}%%%

\paragraph{Validation of widespread refactoring tools.}

Proving the correctness of refactoring operations made us point some situations that require a deep understanding of C mechanisms.
For instance, the fact that any library has a direct access to all global variables of the program is unexpected by some programmers.
That experience can be used to review existing refactoring tools for C. 

\paragraph{Refactoring ill-formed programs.}
Some refactoring tools can perform correct transformations in ill-formed programs, for example when there is a syntax error in a part of the program that is not concerned with a local change. 
Ensuring behavior preservation in presence of errors is not easy because one must ensure that the part of the program which has an error is really not impacted by the change.

\paragraph{False-negatives.} The characterization of the set of programs that make our renaming fail while it could be performed without changing the semantics, as for the problem coming  from CompCert block variables as discussed in Sec.~\ref{limit:block-variables}, is difficult because it cannot be done within the CompCert formalization itself.

\paragraph{Preserve the layout, preserve the pre-processing directives.} 
Most refactoring tools preserve the layout of programs and our tool could probably adapted with the techniques they implement.
Preserving pre-processing directives is probably more difficult. This is studied in several papers~\cite{Spinellis:2003, Vittek:2003, Garrido:2005, ParsingC:2009, Cscout:2010}.

\paragraph{Separate compilation.} 
A little engineering effort is required to take separate
compilation into account, and in particular the use of
compiled object files or libraries in projects.
Standard tools like the Unix command \texttt{nm} can probably be used to check that libraries do not use the renamed variable and its new name.

We also have to take inline assembly code into account to complete the tool.

\paragraph{Other refactoring operations.}
Of course, a large set of popular refactoring operations either atomic or composed are waiting to be verified.

Some aspects of our proof rely on some characteristics of the renaming operation, such as its invertibility, or the fact that it does not change the control flow of programs.
So we expect that some parts of the proofs will change for
other basic (atomic) refactoring operations.
However, other aspects of our proof, such as relaxed preservation, or the correspondence between execution states, can be easily reused.

A large number of interesting refactoring operations are composite: they are combinations of basic operations. 
One of our goal for future work is to be able to prove the correctness of
composite operations and the generation of their preconditions~\cite{composition-of-refactorings2004, Cohen:2013} and to apply it
 to large transformation we have described in~\cite{Cohen:2012} and~\cite{Ajouli:2013}. 

\section*{Aknowledgements}
The author is grateful to Sandrine Blazy, University of Rennes 1, for her help with the use of CompCert.

\bibliographystyle{eptcs} 
\bibliography{biblio}

\appendix

\section{Non-trivial aspects of the proof}%%%%%%%%%%%%%%%%%%%%%%%
\label{sec:proof-explaination}

We report here two specific aspects of the proof of correctness that deserve to be discussed.

\subsection{Higher Order Contexts} 

The concept of context is  familiar in programming language semantics. It is  used to specify places where computations can occur.
In CompCert,
contexts are represented by native Coq functions (type $expr\rightarrow expr$), adopting the higher-order abstract syntax style~\cite{Pfenning:1988}.
Higher order data-structures are generally used because they allow to reuse the mechanisms of function application of the host language instead of re-encoding it. But they also have the well-known drawback of being difficult to inspect and transform. So, the implementation of a renaming in contexts is not trivial.

To transform contexts, we first flatten them by applying them to a fresh witness variable (type $expr$), which gives a plain expression.
Then we transform that expression (normal renaming in an expression),
 and then we build a function by embedding a substitution mechanism where the witness variable has the role of the placeholder for the formal parameter. See the details in our source code (module \texttt{Contexts}).

As a result, any reasoning on context transformations in the Coq development becomes unnatural whereas most proofs on plain expressions are close to the way you would do it ``on paper''.

\subsection{Bindings in Continuations} 

Continuations are another familiar concept in programming language semantics.
CompCert uses them to define the small-step semantics of C.
Here is an extract of the datatype for continuations in CompCert:

\begin{verbatim}
Inductive cont: Type :=
  | Kstop: cont
  | Kseq: statement -> cont -> cont 
  | ...
  | Kreturn: cont -> cont        
  | Kcall: function -> env -> (expr -> expr) -> type -> cont -> cont.
\end{verbatim}

All the constructors in this datatype are linear (they take at most one continuation as parameter). 
This makes continuations ``homomorphic'' with lists. 

When propagating a renaming in a continuation, dealing with
bindings and shadowings is more subtle than bindings in functions. Indeed,
the binders are not explicit as they are in functions.
The only way to bind parameters in continuations is in the functions appearing as the first parameter of the \texttt{Kcall} constructor.
The scope of that binding is the ``segment'' of that continuation which begins with the given \texttt{Kcall} constructor and finishes with the next \texttt{Kcall}, or, if the opening \texttt{Kcall} was the last one of the continuation, at the end of the continuation (\texttt{Kstop}).

Moreover, the first segment of a continuation may not begin with a \texttt{Kcall} constructor.
In this situation, the continuation does not contain enough information to determine the bindings in that segment.
That information has to be found outside of that continuation: in the \texttt{state} construction that embeds that continuation.

It is essential to take all this into account to correctly construct the correspondence \linebreak \mbox{\texttt{State.change\_ident}} of Fig.~\ref{fig:step-commut}.

\end{document}